\documentstyle[aps]{revtex}

\def\deriv#1#2{
{{d #1} \over {d #2}}
}

\def\pderiv#1#2{
{{\partial #1} \over {\partial #2}}
}

\begin{document}
\draft

\title{On the r\^ole of mass growth in dusty plasma kinetics}

\author{A. M. Ignatov \cite{email}}
\address{General Physics Institute, \\ 38, Vavilova str., 117942, \\ Moscow, Russia}

\author{S. A. Trigger \cite{also}}
\address{Institute for High Temperatures, \\ 13/19, Izhorskaya str., 127412 \\ Moscow, Russia}

\maketitle

\begin{abstract}
It is shown that absorption of ions by the dust grain may reduce
the effective translational temperature of the dust component
below the ambient gas temperature.

\end{abstract}

\pacs{PACS number(s): 52.25.Zb, 52.25.Dg}

\section{Introduction}

   The kinetic theory of dusty plasmas, which takes into account specific processes
    of charging, has been considered on phenomenological basis and used for different
     applications in \cite{1,2}. The form of charging collision integrals considered
      in these papers has been recently rigorously justified in \cite{3,4}, where also
       the stationary solution of kinetic equation for charge and velocity
       distributions of grains were established. As it has been established in
       \cite{3}
        the process of absorption of the small particles by grains in dusty plasmas
        can lead to inequality of the grain temperature and the temperatures of the
         light components (even for the case of equal temperatures of electrons and ions).

   At the same time the preliminary results of
    MD simulations for the kinetic energy of dust  particles
    \cite{5}
     demonstrated that mass transfer from light components to dust
      can be essential for the large time scale.
     The problem of mass transfer is also very actual for the conditions
      of the experiments with growth of grains \cite{6} and formation of new materials
      \cite{7}
      in dusty plasmas, as well as for many other applications.
      Therefore the appropriate kinetic theory should be developed, in
      which a new kinetic variable --- the mass of grains ---
must be introduced \cite{8}.

   In this paper we will consider kinetics of the dust particles with variable mass,
    which increases due to absorption of the ambient plasma
     and determine the nonstationary distribution function for
      the grains. It is shown that asymptotically  the effective
       temperature of the dust component is lower than
        the stationary
temperature of the gas. It means the mass growth
 leads to the cooling of the dust component.

\section{Kinetic equations}

Since our main purpose here is to demonstrate the importance of
the mass growth, we simplify the problem ignoring the grain
charge. In other words, we treat a plasma as a neutral gas. We
adopt here that in process of an elementary collision
 a grain absorbs every  atom hitting its surface. An atom
transfers its momentum to a grain and, respectively, the mass of a
grain changes. Therefore generally the distribution
 function
of the dust component depends both on grain momenta, $\bf P$, and
masses, $M$.

 It should noted that the assumption of the complete
absorption seems justified under the conditions of the experiments
aimed at the plasma synthesis of fine grains \cite{6,7}.
Otherwise, an ion hitting the grain surface rather leaves it as a
neutral atom carrying away some momentum. The latter may be
ignored if the surface temperature of a grain is,  by some means,
below the ion temperature.

 The appropriate kinetic equation describing the
process may be written as

\begin{eqnarray}
&&\deriv{f_d({\bf P},M,t)}t =  I_d({\bf P},M,t)= \hfill
\nonumber\\
 && \int d{\bf p} f_n({\bf p}) \left\{  w({\bf p} ,
{\bf P- p} , M-m) f_d({\bf P-p},M-m) -  w({\bf p} , {\bf P} , M)
 f_d({\bf P},M) \right\},  \label{dust}
\end{eqnarray}
where  $  f_n({\bf p})$ is the distribution function of  neutral
atoms of the mass $m$.  The probability of absorption is given by

\begin{equation}
w({\bf p} , {\bf P} , M)=\sigma(M) \left| \frac{\bf P}{M} -\frac{\bf p}{m} \right|, \label{prob}
\end{equation}
where the cross-section, $\sigma(M)$, generally is mass dependent.
For example, assuming the permanent specific gravity of the grain
material results in $\sigma(M) \propto M^{2/3}$.  The distribution
 function in
Eq.~(\ref{dust}) is normalized to the average density:

$$ n_d=\int d{\bf P} d M f_d({\bf P},M). $$

The evolution of the neutral gas distribution is governed by

\begin{equation}
\deriv{f_n({\bf p})}{t}= - \int d{\bf P} dM
   w({\bf p} , {\bf P},M)  f_d({\bf P},M) f_n({\bf p}).
   \label{neutral}
   \end{equation}

   Evidently, the set of kinetic equations
(\ref{dust},\ref{neutral}) provides the conservation of the net
number of dust grains and the total momentum. The total energy is
no longer conserving quantity. The physical reason for this is
fairly obvious: a part of kinetic energy of a colliding  atom is
transferred to the kinetic energy of a dust grain, while the
remainder is spend for the heating of the grain surface. The
latter part of the energy balance is out of our consideration.

The collision term in Eq.~(\ref{dust}) is greatly simplified by
expanding it in powers of a small $\epsilon=m/M$ ratio.
Straightforward expansion of Eq.~(\ref{dust}) results in

\begin{equation}
I_d({\bf P},M) = \pderiv{}{P_i} \left[ -\beta_i g({\bf P},M)
+\lambda_{ij} P_j g({\bf P},M)  +\pderiv{}{P_j}\left( \pi_{ij}
g({\bf P},M) \right) \right] - \pderiv{}{M}
 \left( j g({\bf P},M) \right),  \label{FP}
 \end{equation}
where $g({\bf P},M)=\sigma(M) f_d({\bf P},M)$. The kinetic
coefficients introduced in Eq.~(\ref{FP}) are expressed in terms
of the  gas distribution:

\begin{eqnarray}
j&=& \int d{\bf p}\; p f_n({\bf p}), \\
 \beta_i&=& \int d{\bf p}\; \frac{p}{m} p_i f_n({\bf p}), \\
 \lambda_{ij}&=& \frac{1}{M} \int d{\bf p}\; \frac{p_i p_j}{p} f_n({\bf
 p}),\\
 \pi_{ij}&=& \frac1{2m} \int d{\bf p}\; p p_i p_j f_n({\bf p})
 \end{eqnarray}

The first term in Eq.~(\ref{FP}) arises due to the possible
anisotrophy of the ambient gas distribution. Formally, this term
is proportional to $\epsilon^{1/2}$. The remaining terms
describing
 diffusion in the phase space and mass growth are of the order
 $\epsilon$.

 With sufficiently small number of dust grains one can ignore the
 deviation of ambient gas from initial distribution. Assuming that
 $f_n({\bf p})$ is given by Maxwellian distribution with the
 temperature, $T_n$, and particle density, $n_n$, we get

\begin{equation}\label{isot}
\deriv{f_d( P,M,t)}t= j_0 \left\{ \frac{g( P,M)}M + \frac{P}{3M}
\pderiv{g(P,M)}P + \frac23 T_n \frac1{P^2} \pderiv{}{P} P^2
\pderiv{g(P,M)}P  -\pderiv{g( P,M)}M \right\},
\end{equation}
where it is also supposed that the dust distribution is isotropic.
The coefficient, $j_0$, in Eq.~(\ref{isot}) is the mass flow at
the grain surface, $j_0= n_n \sqrt{8 T_n/\pi}$.

It should be noted that with the last term in the right-hand side
of Eq.~(\ref{isot}) omitted, i.e., in neglecting the process of
the mass growth, there is an exact stationary solution to
Eq.~(\ref{isot}) in the form of Maxwellian function with the
temperature $T_d=2 T_n$. The same conclusion stems also from the
more general approach of \cite{3}.

\section{Temperature evolution}

With the help of Eq.~(\ref{isot}) we study the evolution of the
temperature of the dust component. Although it is possible to
obtain the general solution to Eq.~(\ref{isot}), the corresponding
expression is rather bulky (see Appendix) and little informative.
To grasp the r\^ole of the mass growth one can neglect the mass
dispersion of the dust component looking for the solution in the
form of

\begin{equation}
f_d(P,M,t)=F(P,t)\delta(M-\mu(t)).
\end{equation}

Substituting this into Eq.~(\ref{isot}) yields

\begin{eqnarray}
\deriv{\mu(t)}t&=&j_0 \sigma(\mu(t)) \label{mg}\\
\pderiv{F(P,t)}t&=& j_0 \sigma(\mu(t)) \left\{
\frac{F(P,t)}{\mu(t)}+\frac{P}{3 \mu(t)} \pderiv{F(P,t)}P +
\frac23 T_n \frac1{P^2} \pderiv{}P P^2 \pderiv{F(P,t)}P \right\}.
\label{diff}
\end{eqnarray}

Eq.~(\ref{mg}) shows that the mass of all grains increases
 with the rate determined by the current value of the
 cross-section. The solution to the second equation (\ref{diff})
 is sought in the form of the Maxwellian distribution

 \begin{equation}
 F(P,t)=\frac{n_d}{( 2 \pi \Delta)^{3/2}} e^{-P^2/2  \Delta}  \label{maxwell}
 \end{equation}
with  the time-varying effective temperature,  $ \Delta=T_{eff}(t)
\mu(t)$. Substituting Eq.~(\ref{maxwell}) to Eq.~(\ref{diff})  we
get

\begin{equation}
\deriv{T_{eff}(t)\mu}t =\frac23 j_0 \sigma(\mu)
 ( 2 T_n - T_{eff}(t)). \label{disp}
\end{equation}

 In neglecting the mass growth, as it was already mentioned, the
stationary state of the dust component is characterized by the
effective temperature twice as the gas temperature, $T_{eff} = 2
T_n$. However, the joint solution of Eqs.~(\ref{mg},\ref{disp})
results in

\begin{equation}
T_{eff}(t)= \frac45 T_n + C \mu(t)^{-5/3}, \label{teff}
\end{equation}
where $C$ is an integration constant. Thus, the mass growth yields
cooling of the dust component below the gas temperature, $T_{eff}
\to \frac45 T_n$.

\section{Conclusions}

   We have considered  the kinetic equation for
    the ensemble of grains imposed in neutral gas.
     The process of gas absorption by grains leads to the time
     dependence
      of grain distribution function  due to the  mass growth of the  dust
       particles. For the Maxwellian distribution of neutral gas we found the
        general nonstationary solution of the kinetic equation with variable mass.
       The
         average kinetic energy of grains, that is, the effective temperature of  the dust
          component, tend  to the stationary values. The process
          of establishing of the effective temperature can be interpreted in this case
           as an effective cooling.

\acknowledgments

This work was performed under the financial support granted by the
Netherlands Organization for Scientific Research (NWO), grant \#
047-008-013. One of as (A.M.I.) also acknowledges the support from
Integration foundation, project \# A0029.

\appendix

\section{}

It is a matter of straightforward substitution to verify that the
general solution to Eq.~(\ref{isot}) with an initial condition
$f_d( P,M,0)=f_0(P,M)$ is given by

\begin{eqnarray}
f_d(P,M,t)=&&\int\limits_0^\infty dP^\prime
 \frac{P^\prime M^{2/3}  \sigma(\mu(M,t))}{P \mu(M,t)^{1/3}
 \sigma(M)}  \frac{1}{\sqrt{\pi \Delta(M,t)}}
  \exp\left( -\frac{P^2 M^{2/3} +P^{\prime 2} \mu(M,t)^{2/3}}{4
  \Delta(M,t)} \right) \nonumber\\
  &&\sinh\left( \frac{P P^\prime M^{1/3} \mu(M,t)^{1/3}}{2
  \Delta(M,t)}  \right)  f_0(P^\prime,\mu(M,t), \label{comp}
  \end{eqnarray}
  where $\mu(M,t)$ is a root of the equation

  \begin{equation}
  \int\limits_{\mu(M,t)}^M \frac{d M^\prime}{\sigma(M^\prime)}
  =j_0 t
  \end{equation}
  and $\Delta(M,t)=\frac25 T_n \left( M^{5/3} -\mu(M,t)^{5/3}
  \right)$. Evaluating the average kinetic energy, $\langle {\bf
  P}^2/2M \rangle$, with the help of Eq.~(\ref{comp})
  one can verify that it tends to $6/5 T_n$ even for an arbitrary mass distribution.


\begin{references}
\bibitem[*]{email} Electronic address: aign@fpl.gpi.ru
\bibitem[**]{also} Also at Eindhoven University of Technology,
 P.O. Box 513, MB 5600 Eindhoven, The Netherlands
\bibitem{1} A.M.Ignatov, Plasma Phys. Rep.,  24 (1998) 677 .
\bibitem{2} S.A. Trigger, P.P.J.M.Schram, J.Phys.{\bf D}: Applied Phys.,  32(1999)  234 .
\bibitem{3} A.G.Zagorodny, P.P.J.M. Schram, S.A.Trigger,
Phys.Rev.Lett. 84 (2000) 3594 .
\bibitem{4} P.P.J.M.Schram, A.G.Sitenko,
S.A.Trigger, A.G.Zagorodny, Phys.Rev. {\bf E},  submitted, May
2000.
\bibitem{5} A.M.Ignatov, S.A.Maiorov, P.P.J.M.Schram, S.A.Trigger, P.N. Lebedev Inst. Rep.,
 in print, 2000.
\bibitem{6} F. Vivet, A. Bouchoule, L. Boufendi, J. Appl. Phys. 83 (1998) 7474
\bibitem{7} E.Stoffels,W.W.Stoffels,G.M.W.Kroesen, F.J. de Hoog,
J.Vac.Sci.Technology,{ \bf A} 14 (1996) 556  .
\bibitem{8} S.A.Trigger, Abstracts and Proceedings of IV European Workshop
on Dusty and      Colloidal Plasmas, 3-5 June 2000, Costa da
Caparica, Portugal.
\end{references}
\end{document}